\documentclass[12pt]{iopart}

\usepackage{graphicx}
\usepackage{bm}
\usepackage{epstopdf}
\usepackage{amsfonts}
\usepackage{latexsym,enumerate}
\usepackage{epsfig}
\usepackage[colorlinks=true,linkcolor=blue,urlcolor=blue,citecolor=blue]{hyperref}
\usepackage[T1]{fontenc}
\usepackage[latin9]{inputenc}
\usepackage{hyperref}
\hypersetup{
    colorlinks,
    citecolor=red,
    filecolor=green,
    linkcolor=blue,
    urlcolor=brown
}
\usepackage{iopams}
\hypersetup{linktocpage}
\usepackage{amssymb}
\usepackage{amsthm}
\usepackage[lofdepth,lotdepth]{subfig}

\newcommand{\ket}[1]{\left\vert#1\right\rangle}

\newcommand{\half}{\mbox{$\textstyle \frac{1}{2}$} }

\theoremstyle{definition}

\def\beq{\begin{equation}}
\def\eeq{\end{equation}}
\def\bearr{\begin{eqnarray}}
\def\eearr{\end{eqnarray}}
\def\beal{\begin{equation}\begin{array}{ll}}
\def\eeal{\end{array}\end{equation}}

\def\cn{{\cal N}}
\def\cs{{\cal S}}
\def\cz{{\cal Z}}
\def\CC{\mathbb{C}}

\begin{document}
\title{Implications of non-Markovian quantum dynamics for the Landauer bound}
\author{Marco Pezzutto$^{1,2,3}$, Mauro Paternostro$^3$, and Yasser Omar$^{1,2}$}
\address{$^1$ Instituto de Telecomunica\c{c}\~{o}es, Physics of Information and Quantum Technologies Group, Lisbon, Portugal}
\address{$^2$ Instituto Superior T\'ecnico, Universidade de Lisboa, Portugal}
\address{$^3$ Centre for Theoretical Atomic, Molecular and Optical
Physics, School of Mathematics and Physics, Queen's University
Belfast BT7 1NN, United Kingdom}
\ead{marco.pezzutto@tecnico.ulisboa.pt}

\begin{abstract}
We study the dynamics of a spin-1/2 particle interacting with a multi-spin environment, modelling the corresponding open system dynamics through a collision-based model. The environmental particles are prepared in individual thermal states, and we investigate the effects of a distribution of temperatures across the spin environment on the evolution of the system, particularly how thermalisation in the long-time limit is affected. 
We study the phenomenology of the heat exchange between system and environment and consider the information-to-energy conversion process, induced by the system-environment interaction and embodied by the Landauer principle. Furthermore, by considering an interacting-particles environment, we tune the dynamics of the system from an explicit Markovian evolution up to a strongly non-Markovian one, investigating the connections between non-Markovianity, the establishment of system-environment correlations, and the breakdown of the validity of Landauer principle. 

\end{abstract}

 
\section{Introduction}
\label{sec:intro}

Non-Markovian open-system dynamics has recently received considerable attention~\cite{Breuer16,deVega15}, including the formulation of figures of merit for its characterization ~\cite{Breuer09,Laine10,Rivas10,Chruscinski11,Chruscinski14, Lorenzo13}, and the first steps towards its experimental assessment~\cite{chiuri,cinesi,cinesi2,brasiliani,recover,brasiliani2}.  While a full understanding of the origins of non-Markovianity~\cite{mazzola,smirne}, and the formulation of a universal characterization of is implications are the objects of current investigations~\cite{fanchini,buscemi,acin,Pollock15,Addis15}, the community interested in open system dynamics has recently recognised the relevance of non-Markovianity for the assessment of the properties of non-equilibrium quantum systems~\cite{Kutvonen14,Bylicka15,Goold15,Guarnieri16}. In particular, the role of memory effects (which are believed to be a key cause of the emergence of non-Markovian effects) in logically irreversible processes has recently attracted some attention~\cite{Lorenzo15,ReebWolf14} in light of the relevance that Landauer principle has for information processing  at both the classical and quantum level~\cite{Landauer61,pleniovitelli,Mohammady16}. The relevance of the principle at the quantum scale has been largely debated, with contributions both at the theoretical~\cite{Goold15,Lorenzo15,ReebWolf14,Anders10,Hilt11,Barnett13,Berut15,Mohammady16} 
and the experimental level~\cite{Berut12,Koski14,Silva16}.

\begin{figure}[t!]
\center{{\bf (a)}\hskip4.5cm{\bf (b)}\hskip4.5cm{\bf (c)}}\\
\begin{center}
{\includegraphics[width=0.26\textwidth]{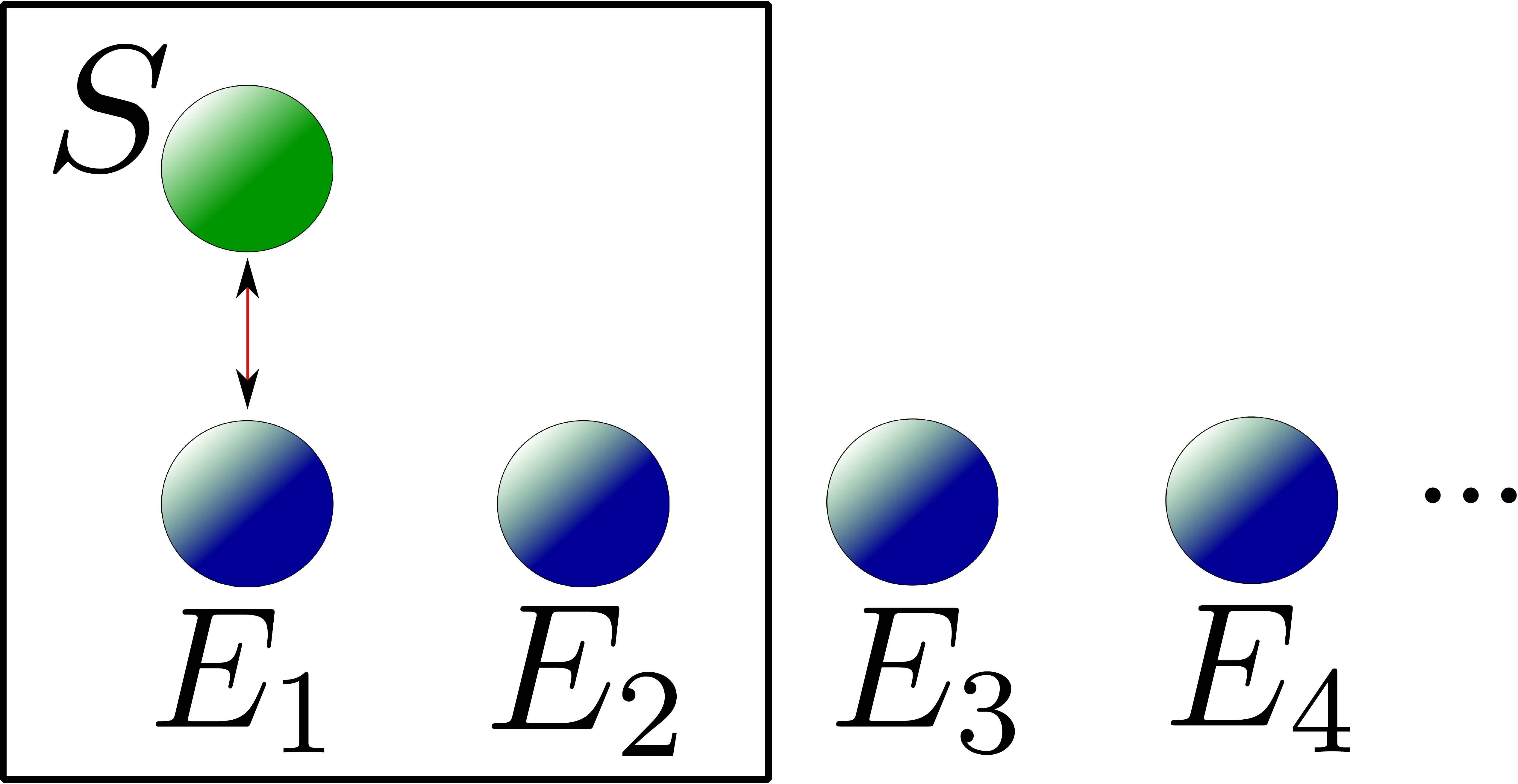}}
 \qquad
{\includegraphics[width=0.26\textwidth]{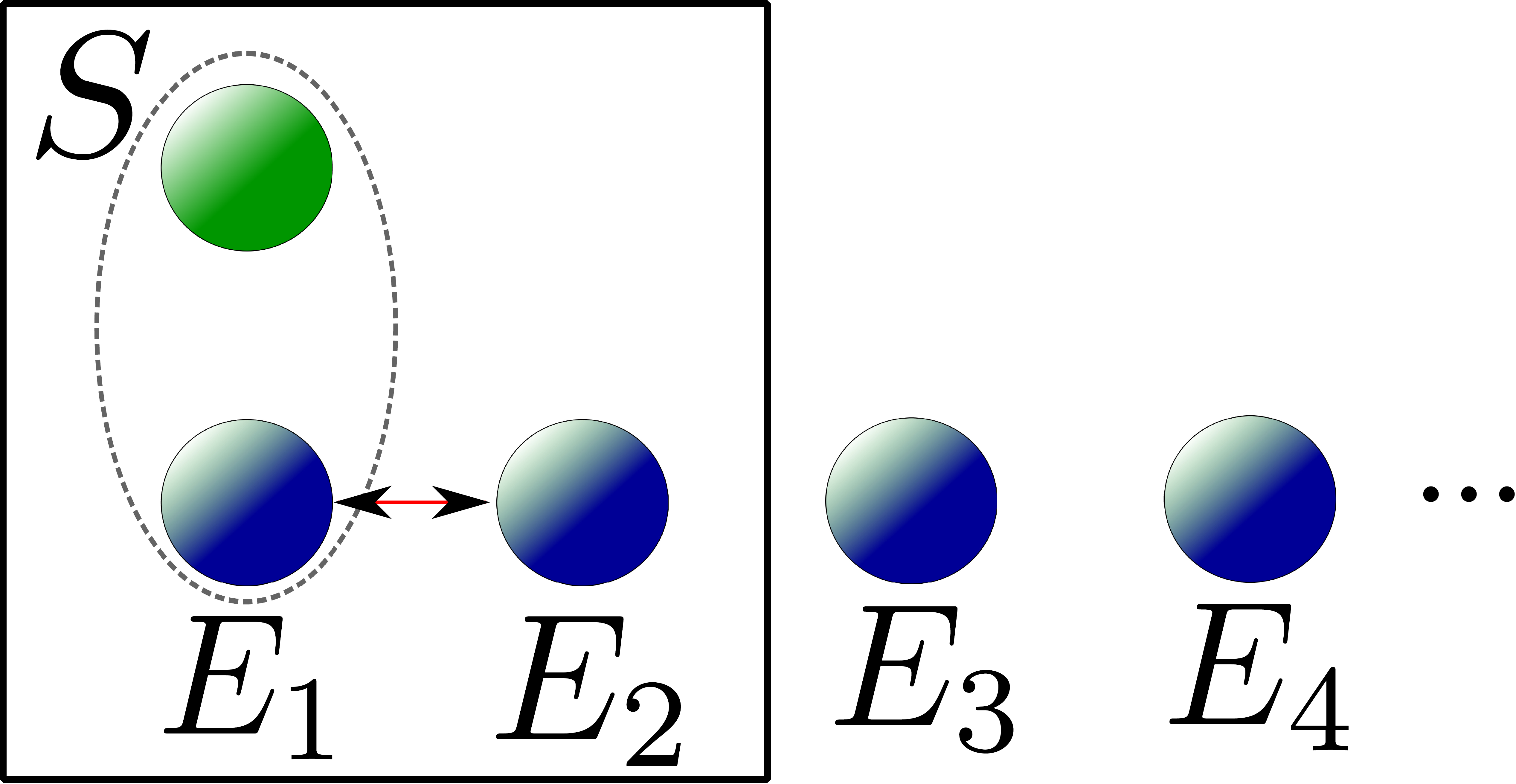}}
 \qquad
{\includegraphics[width=0.26\textwidth]{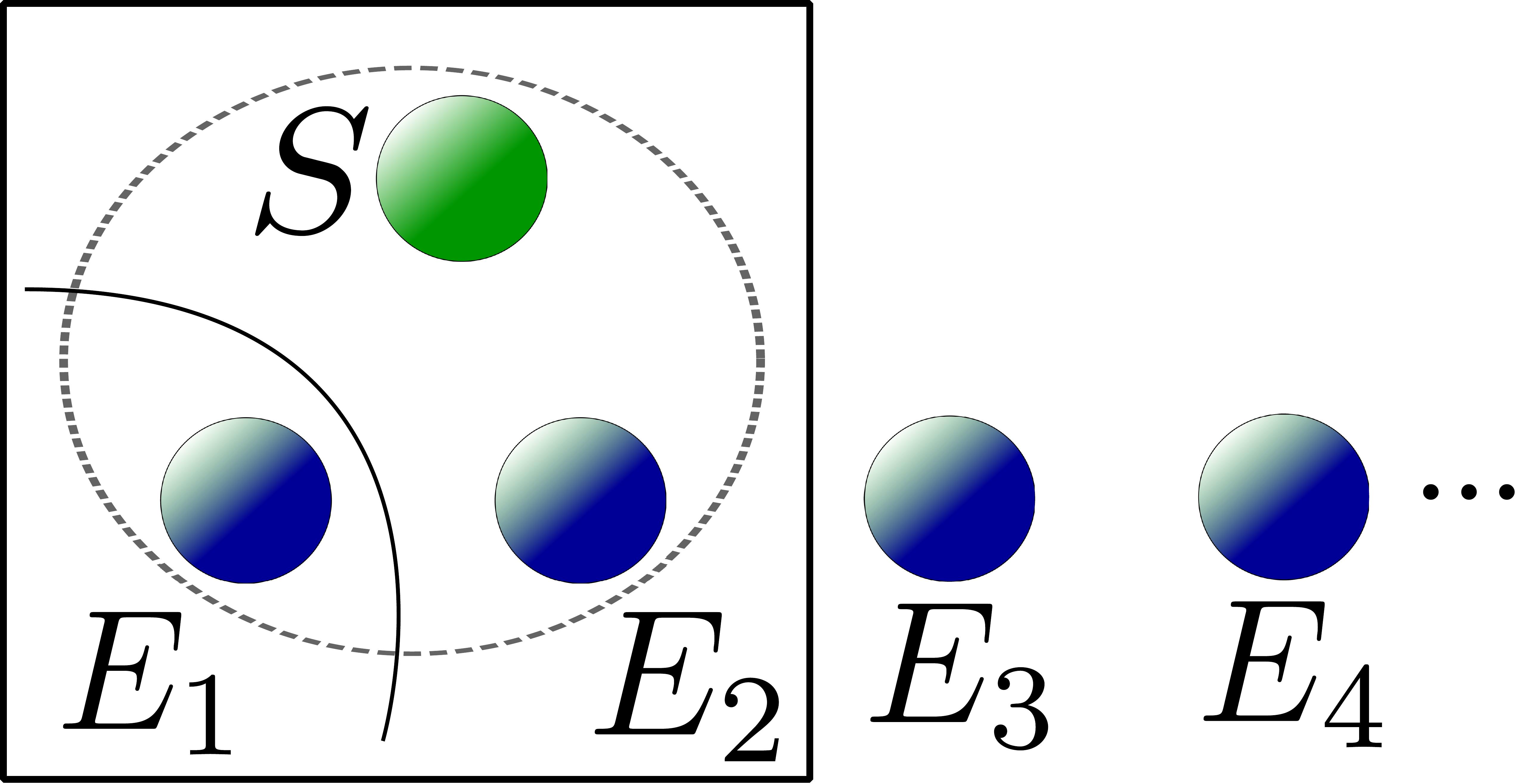}}
\end{center}
\center{{\bf (d)}\hskip4.5cm{\bf (e)}\hskip4.5cm{\bf (f)}}\\
\begin{center}
{\includegraphics[width=0.26\textwidth]{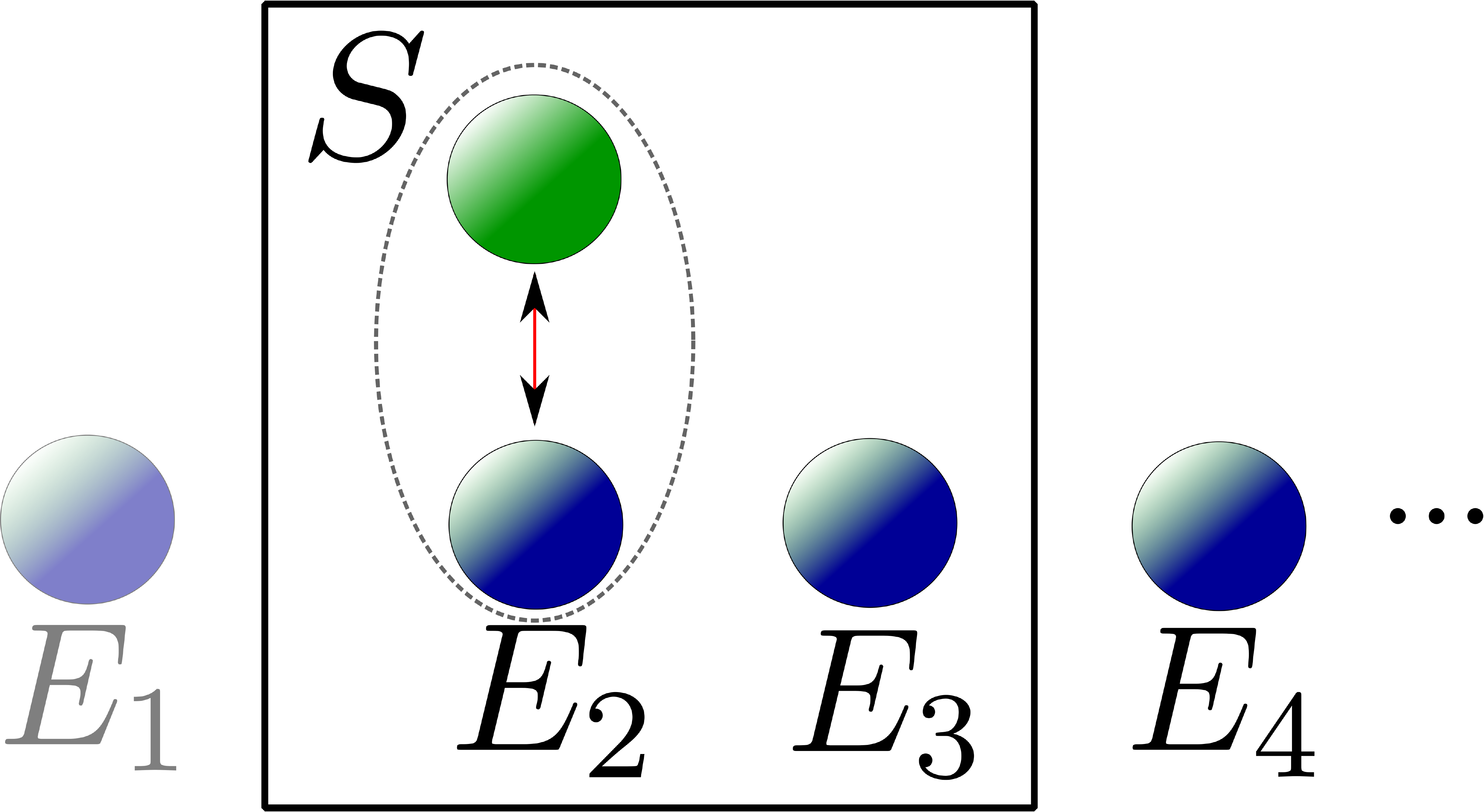}}
\qquad
{\includegraphics[width=0.26\textwidth]{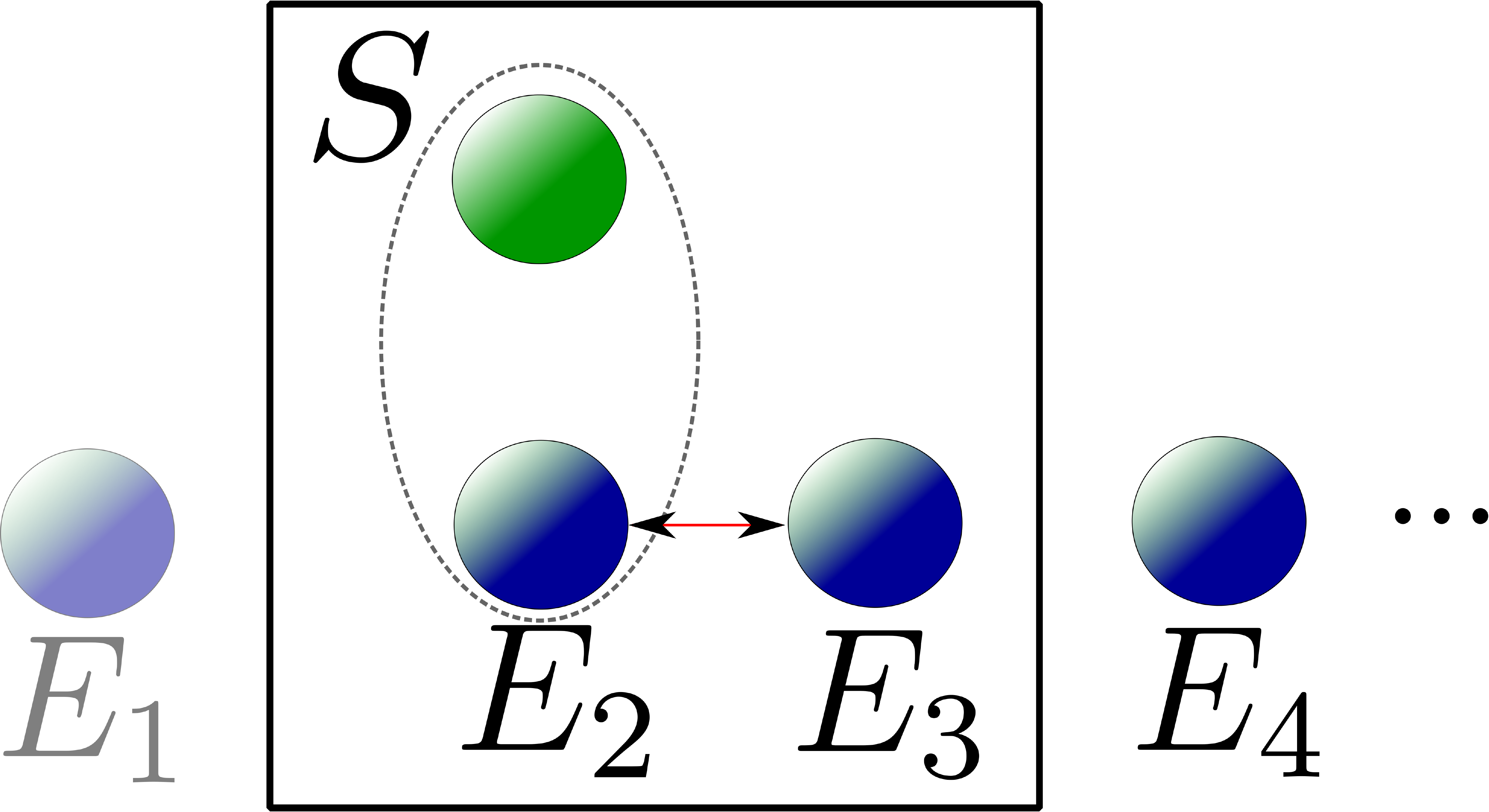}}
 \qquad
{\includegraphics[width=0.26\textwidth]{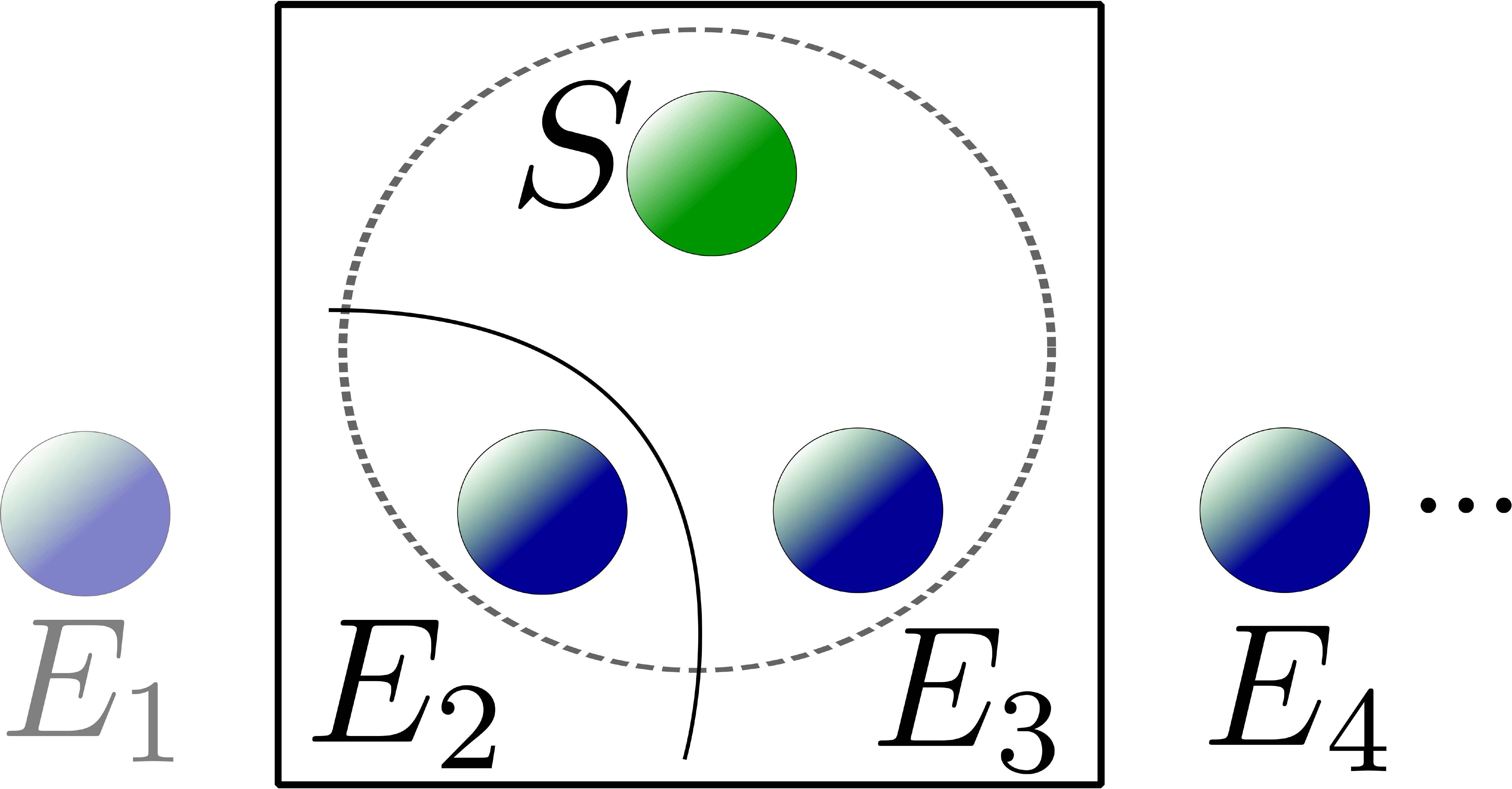}}
\end{center}
\caption{\textbf{Non-Markovian dynamics.} We study the dynamics of a spin-1/2 particle, the system $\cs$, undergoing a sequence of interactions with similar spin-1/2 environment particles. At each step we work with a dynamical cell composed by the system and two environment particles. Here we sketch the first two iterations of the protocol. The boxes represent the dynamical cell, double arrows represent collisions through Heisenberg interaction and dashed contours are correlations. {\bf (a)} The dynamical cell contains $\cs$ and particles $E_1$ and $E_2$, all initially uncorrelated. Here $\cs$ interacts with $E_1$. {\bf (b)} The system and $E_1$ are now correlated; $E_1$ interacts with $E_2$. {\bf (c)} All the three particles in the cell are correlated; $E_1$ is traced away and removed from the cell. {\bf (d)} The process is iterated: the cell contains now $\cs$, $E_2$ and $E_3$ and here $\cs$ collides with $E_2$. Correlations between the two are carried over from the previous step. {\bf (e)} Collision $E_2$-$E_3$. {\bf (f)} Particle $E_2$ is traced away; to iterate the dynamics, $E_4$ will then be added to the cell.
\label{fig:non-MK-dynamics}}
\end{figure}

In this work, we shed further light on the interplay between environmental memory effects and logical irreversibility in non-equilibrium processes. We construct a collisional model of the open-system dynamics of a spin-1/2 particle, of which figure~\ref{fig:non-MK-dynamics} is an example, using and extending significantly a framework that was also used recently to investigate non-Markovianity~\cite{McCloskey14,Kretschmer16}. Our model consists of a sequence of discrete-time collisions, each ruled by a Heisenberg Hamiltonian, between the system and one environmental particle at a time. For non-interacting environmental particles all prepared in the same state, the system undergoes a homogenization dynamics~\cite{Scarani02}: in the long-time limit corresponding to a large number of collisions, the system's state is asymptotically driven towards the initial preparation of the environmental particles. Such homogenization is relatively robust against state fluctuations across the multi-particle environment induced, for instance, by a spatially inhomogeneous temperature. 

For an environment of interacting particles, instead, the effective dynamics that the system undergoes can be tuned broadly from a fully Markovian to a highly non-Markovian regime. The crossover from Markovianity to non-Markovianity has long been object of investigation, both experimentally \cite{Rodriguez08} and theoretically \cite{GonzalesTudela10}, with particular attention to non-Markovian effects due to a hierarchical environment \cite{Ma14,Man15a}, the role of a non-Markovian memory-keeping environment in harnessing the quantum memory stored in a qubit \cite{Man15b,Man15c} and the ability to control the (non-)Markovian character of the dynamics of a composite system by tuning the internal couplings \cite{Brito15,LoFranco15}. 

In our study case, collisions occurring between environmental subsystems have a twofold effect: on the one hand, they induce system-environment correlations resulting in memory effects that allow the environment to retain information on the state of the system at previous steps of its discrete-time evolution; on the other hand, they enable a feedback process whereby information is coherently brought back into the state of the system, thus steering its state in a distinguished way with respect to the corresponding homogenization dynamics. 
We unveil the implications that such a rich dynamics has for logical irreversibility, assessing the break down of Landauer principle~\cite{Landauer61} as the non-Markovian character of the system's evolution is enhanced. In particular, we show a causal link between the system-environment correlations and the opening up of temporal {\it windows} in the time evolution of the system within which Landauer bound is no longer valid.

A connection between correlation revivals and non-Markovianity was also reported in \cite{recover,Xu13,DArrigo14}, although in these works the authors investigate the dynamics of correlations within a composite system interacting with a local environment. Furthermore, it is worth mentioning that a breakdown of the II Law of Thermodynamics in the form of the Clausius inequality due to system-environment quantum correlations has also been evidenced in harmonic systems coupled to a bath of harmonic baths \cite{Allahverdyan00,Allahverdyan02,Horhammer05}, and that in the context of generalizations of the II Law for systems strongly coupled to a reservoir,  \cite{Hilt11,Williams11,Seifert16}, the observed violations are only apparent.

The remainder of this work is structured as follows: in section~\ref{sec:non-markovianity-theory} we summarize some key results on quantum non-Markovianity; in section~\ref{sec:model} we introduce our collision model, its thermodynamics and the formulation of the Landauer principle for non-equilibrium quantum systems. Section~\ref{sec:res-non-int} presents the results obtained with the fully Markovian dynamics, in particular the emergence of homogenization, while in section~\ref{sec:res-nmk} we investigate the non-Markovian regime and show how non-Markovianity, local violations of the Landauer principle and correlations are causally connected. Section~\ref{conc} draws our conclusions and presents some open questions that will be addressed in future works.

\section{Quantum non-Markovianity}
\label{sec:non-markovianity-theory}

\subsection{Definition and measure}

The approach we use to define and measure quantum non-Markovianity, from~\cite{Breuer09}, employs the \emph{trace distance}~\cite{NielsenChuang10} between two quantum states $\rho_1$ and $\rho_2$
\begin{equation}
D(\rho_1,\rho_2) := \half \| \rho_1 - \rho_2 \| \,, \qquad \| \rho \| = \Tr \sqrt{\rho^{\dag} \rho} \,,
\end{equation}
which is a metric in the space of density matrices. Two properties are particularly relevant: a) the trace distance is a measure of the \emph{distinguishability}
between states and b) it is \emph{contractive} under positive trace preserving quantum dynamical maps, even if not completely positive. 
Let then ${\{} \Phi_t {\}}$ be a family of quantum dynamical maps, $\rho_1$ and $\rho_2$ two initial states and $\rho_{1(2)}(t) \equiv \Phi_t \rho_{1(2)}$ the corresponding evolved states. The dynamics given by the process $\Phi$ is \emph{Markovian} if, for any pair of initial states, the trace distance $D(\rho_1(t),\rho_2(t)) $ decreases monotonically for all $t\ge 0$.
Conversely, a quantum dynamical process $\Phi$ is said to be \emph{non-Markovian} if there exists a pair of initial states for which the trace distance between the evolved states is increasing in some time intervals. In other words, there exists two states $\rho_1$ and $\rho_2$ and some time $t \ge 0$ at which $ \partial_t D\big( \rho_1(t), \rho_2(t) \big)$ is strictly positive. 
A non-Markovian process can thus increase the distinguishability between two initially different states: the environment has some \emph{memory effect} on the system dynamics. 
This fact can be used to quantify non-Markovianity.
Denoting $\sigma(t):= \partial_t D(\rho_1(t), \rho_2(t) )$, the \emph{degree of non-Markovianity} $\cn$ of the quantum dynamical process $\Phi$ is
\beq
\label{eq:N}
\cn(\Phi) := \max_{ {\{ } \rho_1,\rho_2 {\}} } \int_{0}^{+\infty}   \frac{1}{2}\Big(|\sigma(t)| + \sigma(t)\Big) \rmd t .
\eeq
The integrand is non-null only in the intervals where the derivative is positive. Then a maximization over all pairs of initial states is performed, and it was shown in~\cite{Wissmann12} that any two states maximising $\cn$ belong to the boundary of the state space and are orthogonal. Finally, as exposed in section~\ref{sec:model}, our dynamics is implemented in discrete time steps, therefore we compute the measure $\cn$ by substituting the derivative $\sigma(t)$ with the difference of the trace distance at steps $n$ and $n-1$, $D(\rho_{1,n},\rho_{2,n}) - D(\rho_{1,n-1},\rho_{2,n-1})$.

\subsection{Non-Markovianity and system-environment correlations}

One important feature of the dynamics is the connection between system-environ\-ment correlations (hereafter referred to simply as {\it correlations}) and the manifestation of non-Markovianity. In~\cite{Mazzola12} the authors provided a link between the behaviour of the derivative $\sigma(t)$ and system-environment correlations quantified by means of the matrix $\chi^{SE}(t):=\rho^{SE}(t)-\rho^{S}(t)\otimes\rho^{E}(t)$, in the form of an upper bound for $\sigma(t)$ depending explicitly on $\chi^{SE}(t)$.
The result connects non-Markovianity with the emergent distinguishability between initially identical environment states, and with the creation of correlations. In that work, the authors provide the example of an $N+1$ spin-1/2 particle system, and find a connection between the time evolution of the trace distance and that of correlations, quantified by the \emph{mutual information} between system and environment. Trace distance and correlations evolve in a periodic and synchronised fashion, both exhibiting a non-monotonic behaviour incompatible with a Markovian dynamics. A similar observation is reported in \cite{McCloskey14}, and in our model we also observe this feature, as shown in section~\ref{sec:res-nmk}, and extend the connection between trace distance and correlations to include also the Landauer bound. In the work \cite{Smirne13} a tighter connection between the trace distance derivative and correlations was provided, in the form of both an upper and a lower bound on $\sigma(t)$.

\section{Dynamical model and thermodynamics}
\label{sec:model}

\subsection{Basic setup and notation}

The system and a generic environment particle are denoted by $\cs$ and $E$ respectively, and described by local spin-1/2 Hamiltonians 
\begin{equation}
H_{\cs(E)} := {\hbar \tilde\omega_{s(\rme)}}\sigma_z/2, \qquad \tilde\omega_{s(\rme)}>0 \, .
\end{equation}
The environmental particles are initially prepared in the thermal state $\rho_{\beta} := e^{-\beta H_E}/\cz$ at temperature $T$, where $\beta=1/\kappa T$ is the inverse temperature and $\cz = \Tr [e^{-\beta H_E}]$ is the corresponding partition function (we use units such that the Boltzmann constant $\kappa$ takes value $1$ from now on). As discussed in section~\ref{sec:res-non-int}, we allow for fluctuations in the preparation of the environmental particles.  The choice of thermal states is consistent with the necessity of a well-defined temperature of the environment. 

The dynamics proceeds through a sequence of interactions, or \emph{collisions}, and we first consider a process in which the system interacts each time with a new environment particle, implementing the idea of a large, memoryless thermal bath. The system-environment interaction is ruled by the Heisenberg Hamiltonian and implemented through the unitary $V_{\cs E}$ 
\beq
\label{eq:VopExp}
H_{int} = \tilde{J}_{\cs E} \Big( \sigma_x^{\cs} \sigma_x^E + \sigma_y^{\cs} \sigma_y^E + \sigma_z^{\cs} \sigma_z^E \Big), \qquad 
V_{\cs E} = \exp \Big[ -\frac{\rmi}{\hbar}  H_{int} \tau\Big],
\eeq
where $\tau$ is the interaction time. We call $H_0 = H_{\cs}+H_E$ the free Hamiltonian of system and environment, so that $U_0 = \exp ( - \frac{\rmi}{\hbar} H_0 \tau_0)$ gives the corresponding free evolution, occurring for a time $\tau_0$ between two consecutive collisions. Starting with the environment particle in the pre-collision state $\rho^E_{pre} \equiv \rho_{\beta}$, the system is brought from step $n$ to step $n+1$ through the process
\beq
\label{eq:SEcollision}
\rho^{\cs}_n \otimes \rho_{pre}^E \quad \mapsto \quad \rho^{\cs E}_{n+1} = U_0 V_{\cs E} (\rho^{\cs}_n \otimes \rho_{pre}^E) V_{\cs E}^{\dag}U_0^{\dag} \, ,
\eeq
ant then the new marginal states $\rho^{\cs}_{n+1} = \Tr_E [ \rho^{\cs E}_{n+1}]$ and $\rho^E_{post} = \Tr_{\cs} [ \rho^{\cs E}_{n+1}]$ are computed.

\subsection{Thermodynamics}

From the marginal states we compute the von Neumann entropy $S_n = -\Tr \big(\rho^{\cs}_n \ln \rho^{\cs}_n)$,\footnote{Consistency between the information theoretical and the thermodynamical contexts bridged by the Landauer principle calls for choosing the natural logarithm, thus quantifying the entropy in \emph{nats}.}
 the variation of the system energy and the exchanged heat at step $n$
\beq
\Delta U_n = \Tr \Big[ H_{\cs} (\rho^{\cs}_n - \rho^{\cs}_{n-1}) \Big], \qquad
\Delta Q_n = \Tr \Big[ H_E (\rho^E_{post}-\rho^E_{pre}) \Big] .
\label{eq:thermo}
\eeq
If $[H_{int},H_0 ]\neq0$ {the interaction does not conserve the total energy of the physical system here at hand}. The difference $W_n:=\Delta Q_n - \Delta U_n$ can thus be interpreted as \emph{work}, either poured into or extracted from the system by the unitary operation $V_{\cs E}$. The case of energy-conserving interactions is realized in the \emph{resonance} condition $\tilde\omega_s=\tilde\omega_\rme$.

We can check the validity of the \emph{Landauer bound} $\beta \Delta Q \ge \Delta S$ at each collision. In~\cite{ReebWolf14}, the principle is given foundation in a framework of quantum statistical mechanics: we let $\Delta S = S_{n-1} - S_n$ be the entropy \emph{decrease} of the system, and we call $\Delta Q$ the heat-exchange introduced in (\ref{eq:thermo}). The bound then holds if:
\begin{enumerate}
\item[{\bf 1)}] No other physical systems is involved;
\item[{\bf 2)}] The environment is initially in a thermal state; 
\item[{\bf 3)}] The system and the environment are not initially correlated;
\item[{\bf 4)}] The dynamics proceeds through a joint unitary evolution. 
\end{enumerate}
We shall see in section~\ref{sec:res-nmk} that inter-environment interactions, implemented as described below, can induce the presence of correlations between the system and an environmental particle before they interact directly, thus contradicting assumption {\bf 3)} and allowing for violations of the bound.
 
\subsection{Implementation of the non-Markovian dynamics}
\label{sec:dyn-cell}
Let us now {\it switch on} the inter-environment interactions, which we alternate to the system-environment ones. 
The discrete-time evolution is achieved thanks to the iteration of a \emph{dynamical cell} comprising the system and two environment particles. We now go through the $n^{\rm th}$ iteration of the scheme, with reference to figure~\ref{fig:non-MK-dynamics} exemplifying the first two steps. At the beginning, the dynamical cell contains $\cs$, $E_n$ and $E_{n+1}$.  The $\cs$-$E_n$ collision occurs via the unitary operation $V_{\cs E}$ in (\ref{eq:SEcollision}). From now on, we absorb the evolution times $\tau$ and $\tau_0$ into the respective rates, so that we now consider dimensionless quantities such as $J_{SE}=\tilde{J}_{\cs E}\tau$, $\omega_{s,{\rm e}}=\tilde\omega_{s,{\rm e}}\tau_0$ and analogous ones.
Then, $E_n$ and $E_{n+1}$ interact through a unitary operation $V_{EE}$ similar to $V_{\cs E}$, where the dimensionless inter-environment coupling constant $J_{EE}$ can be different from $J_{\cs E}$. The three particles in the dynamical cell can now be all correlated. The updated marginal states are computed, and thus the thermodynamical quantities.  In particular, at each step we compute the marginal state of the environment particles before the $\cs$-$E_n$ and after the $E_n$-$E_{n+1}$ interaction, and thus the exchanged heat $ \Delta Q_n = \Tr [ (H_E \otimes H_E) (\rho_{E_nE_{n+1}}^{post} - \rho_{E_nE_{n+1}}^{pre} )]$. In order to iterate the dynamics, $E_n$ is now traced out, we compute the two-particle marginal state $\rho^{post}_{\cs E_{n+1}}$ and a fresh environmental particle $E_{n+2}$, prepared in the thermal state $\rho_{\beta}$, comes into the cell:
\beq
\rho^{post}_{\cs E_nE_{n+1}} \, \to  \,\rho_{\cs E_{n+1}}^{post}\otimes \rho_{\beta} \equiv \rho^{pre}_{\cs E_{n+1}E_{n+2}} \,.
\eeq
The whole process is then iterated.

One of our goals is to investigate the features of the dynamics from a fully Markovian to a completely non-Markovian regime, for which it is useful to express the operator $V_{\cs E}$ in the form of a {partial Swap}, using the following result from \cite{Scarani02} 
\[
e^{-\rmi \frac{\phi}{2}} \exp \Big[ \rmi \frac{\phi}{2}  (\sigma_x \otimes \sigma_x +\sigma_y \otimes \sigma_y + \sigma_z \otimes \sigma_z)\Big] = 
e^{-\rmi \phi} \big( \cos \phi \,\mathbb{I}_4 + \rmi \sin \phi \,U_{sw}  \big), 
\]
where $U_{sw}$ is the two-particle Swap operator: $U_{sw}(\ket{\psi_1} \otimes \ket{\psi_2}) = \ket{\psi_2} \otimes \ket{\psi_1}$ for all $\ket{\psi_1},\ket{\psi_2}$ in $\CC^2$. The partial Swap acts on any 2-particle state by leaving it unchanged with probability $\cos^2 \phi$ and swapping it with probability $\sin^2 \phi$. 
We can now write 
\begin{equation}
V_{\cs E}=e^{-\frac{\rmi}{\hbar} J_{\cs E}} [\cos(2 J_{\cs E}) \,\mathbb{I}_4 + \rmi \sin (2 J_{\cs E}) \,U_{sw} ] \, ,
\end{equation}
and tune the operator from no interaction ($J_{\cs E}=0$) up to a complete Swap ($J_{\cs E}=\pi/4$).

\section{Observation of homogenization in the Markovian dynamics}
\label{sec:res-non-int}
\begin{figure}[t]
{\includegraphics[width=0.40\textwidth]{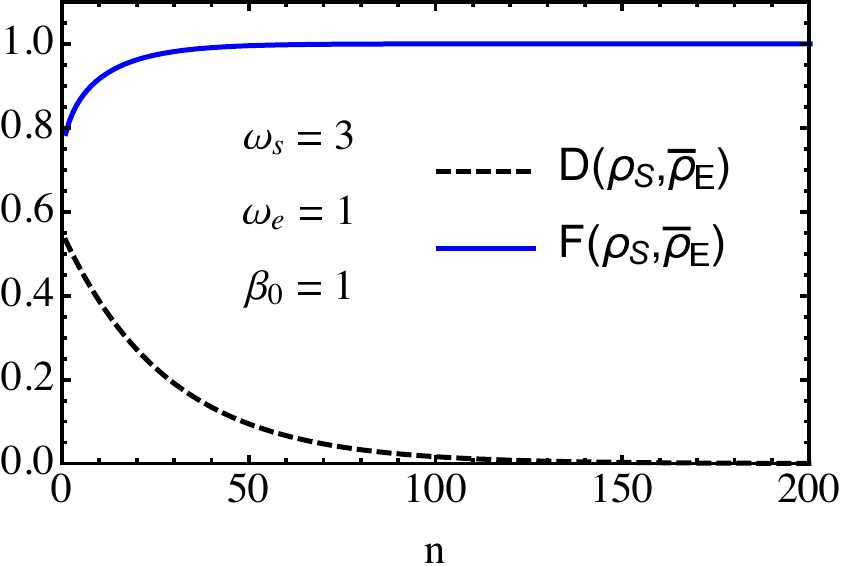}}
\quad\quad\quad
{\includegraphics[width=0.44\textwidth]{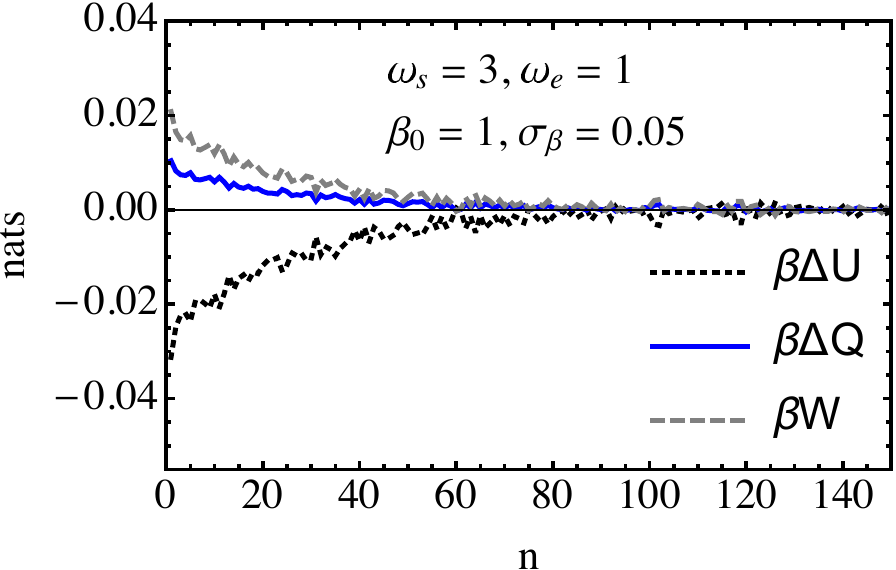}}
\caption{\textbf{Markovian process.} Here we show the main features of the system dynamics in interaction with a Markovian environment.  In both panels, the horizontal axis shows the number of collisions $n$. All plotted quantities are dimensionless. {\bf (a)} Homogenization witnessed by the monotonic behaviour of the distance $D(\rho_{\cs},\bar{\rho}_E)$ and the fidelity $F(\rho_{\cs},\bar{\rho}_E)$ to the average environment state $\bar{\rho}_E$. {\bf (b)} Evolution of the system's (dimensionless) energy change $\beta \Delta U$, the exchanged heat with the environment  $\beta \Delta Q$, and the work $\beta W$ in the presence of noise. Environment particles are prepared in thermal states with inverse temperature $\beta$ chosen from a Gaussian $(\beta_0,\sigma_{\beta})$ distribution. We take the initial system state $\ket{+}_S$ and the coupling constant $J_{\cs E}=\pi/32$, to guarantee conditions of weak $\cs$-$E$ coupling. The system and environment proper frequencies are $\omega_{\cs}=3 $ and $\omega_{E}=1$ respectively. }
\label{fig:homogenizationthermodynamics}
\end{figure}
\begin{figure}[t]
\center{{\bf (a)}\hskip8cm{\bf (b)}}\\
{\includegraphics[width=0.43\textwidth]{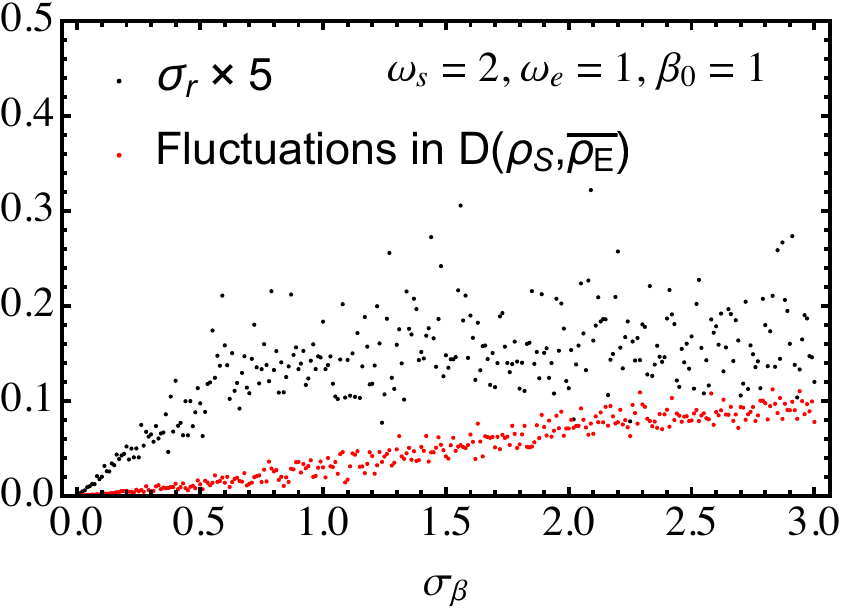}}
\quad\quad
{\includegraphics[width=0.48\textwidth]{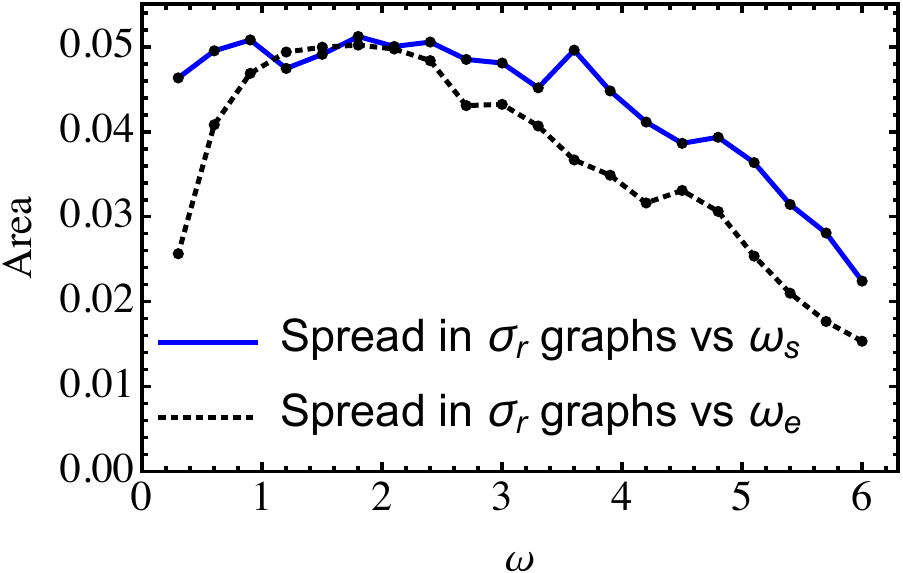}}
\caption{\textbf{Breaking down of homogenization}. {\bf (a)} Asymptotic fluctuations $\sigma_r$ of the system's Bloch vector (black/darker points) and trace distance $D(\rho_{\cs},\bar{\rho}_E)$ from the average environment state (red/lighter points), plotted against the fluctuation $\sigma_{\beta}$ in the environmental state. {\bf (b)} Each point is achieved by estimating the area of the distribution of points in $\sigma_r$-versus-$\sigma_{\beta}$ graphs similar to the example given by the black dots in panel {\bf (a)}. All quantities plotted in dimensionless units.}
\label{fig:sigma-sigmasummary}
\end{figure}
Here we expose the results produced by the dynamics with no inter-environment interactions. If the environmental particles are all prepared in the same state $\rho_{\beta}$, the dynamics produces \emph{homogenization}:  the system reaches asymptotically the very same state in which the environment particles are prepared, $\rho_{\cs} \to \rho_{\beta}$ as illustrated in figure~\ref{fig:homogenizationthermodynamics} {\bf (a)}.\footnote{
Indeed, the equilibration condition $[H_{int},\rho_{\cs} \otimes \rho_{\beta}]=0$ is satisfied only if $\rho_{\cs} = \rho_{\beta}$.}
This could appear to be counterintuitive at first, as one would expect the system to thermalize to the state $\exp[-\beta H_{\cs}]/\cz_{\cs}$. However, the dynamics is effectively governed by a global time-dependent Hamiltonian, and the system experiences an active \emph{driving}. Homogenization occurs even for small fluctuations of the environmental states, as shown in figure~\ref{fig:homogenizationthermodynamics} {\bf (b)}. For each collision, we take $\beta$ from a Gaussian distribution, centred in $\beta_0$ and with amplitude $\sigma_{\beta}$. We call $\bar{\rho}_E$ the average environmental state. The behaviour of the asymptotic fluctuations in the system appears to depend strongly on the entity of those occurring at the environment level. For small fluctuations in the environmental states, the asymptotic fluctuations increase almost linearly. However, for larger values of $\sigma_{\beta}$ a less regular and more chaotic behaviour emerges, as shown in figure~\ref{fig:sigma-sigmasummary} {\bf (a)}. Moreover, the quantitative trend followed by the system's fluctuations appear to depend strongly on the energy spacings $\omega_{s}$ and $\omega_{\rm e}$. In order to characterize such dependence and change of trend more quantitatively, we have estimated, through a Monte-Carlo approach, the area over which the cloud of points in figure~\ref{fig:sigma-sigmasummary} {\bf (a)} is distributed [cf. figure~\ref{fig:sigma-sigmasummary} {\bf (b)}]. This provides an estimate of the {\it spread} of the distribution in the different dynamical situations that we have addressed. For small frequencies, the spread depends more heavily on $\omega_{\rm e}$ than on $\omega_{s}$, while for high frequencies it decreases in both cases. This effect is linked to the fact that, in such conditions, the interaction Hamiltonian becomes weaker than the free Hamiltonian. To summarise, the repeated interaction of the system with a Markovian environment produces homogenization, at least as far as the noise level in the environment is contained. For increasing noise, however, homogenization is lost and the asymptotic dynamics becomes less predictable.

\section{Non-Markovianity, correlations and the Landauer bound}
\label{sec:res-nmk}

Here we present the results for the dynamics with interacting environment, up to the complete Swap between environmental particles. This case is equivalent to having the system interacting repeatedly with one single environment particle. The asymptotic behaviour is the same as in the Markovian case, as non-Markovianity seems to affect only the transient before equilibration. The full Swap case embodies an exception: the system does not homogenize and the dynamics exhibits a periodic trend that is repeated indefinitely. 
\begin{figure}[b]
\begin{center}
\includegraphics{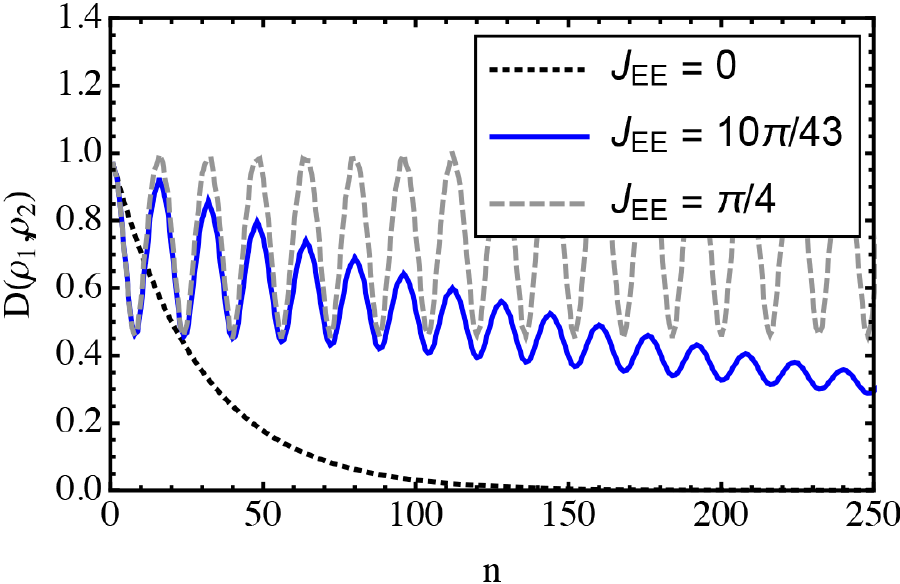}
\end{center}
\caption{\textbf{Non-Markovianity} witnessed by non-monotonicity of the trace distance $D(\rho_1,\rho_2)$ between two evolved states, against the number of collisions $n$. A purely Markovian dynamics causes $D(\rho_1,\rho_2)$ to decrease monotonously, while any increase of it can only be caused by non-Markovianity in the dynamics. Our model allows for this when inter-environment interactions are present.
 We chose $\ket{+}_S$ and $\ket{-}_S$ as the two initial states, which maximise the degree of non-Markovianity $\cn$ \ref{eq:N}. For the black, dotted curve we have taken the inter-environment coupling constant $J_{EE}=0$, corresponding to Markovian evolution; for the blue, continuous one we have $J_{EE}=10\pi/43$, which results in an intermediate case. Finally, the gray, dashed curve is for $J_{EE}=\pi/4$, i.e. a complete Swap, and thus strongly non-Markovian dynamics. The $\cs$-$E$ coupling constant is $J_{\cs E}=\pi/{32}$ for weak coupling, then $\beta =1$ and the system and environment proper frequencies are $\omega_s = 3$ and $\omega_e =1$ respectively.}
\label{fig:trace-dist}
\end{figure}
\begin{figure}[t!]
\center{{\bf (a)}\hskip7.5cm{\bf (b)}}\\
\vskip0.1cm
{\includegraphics[width=0.49\textwidth]{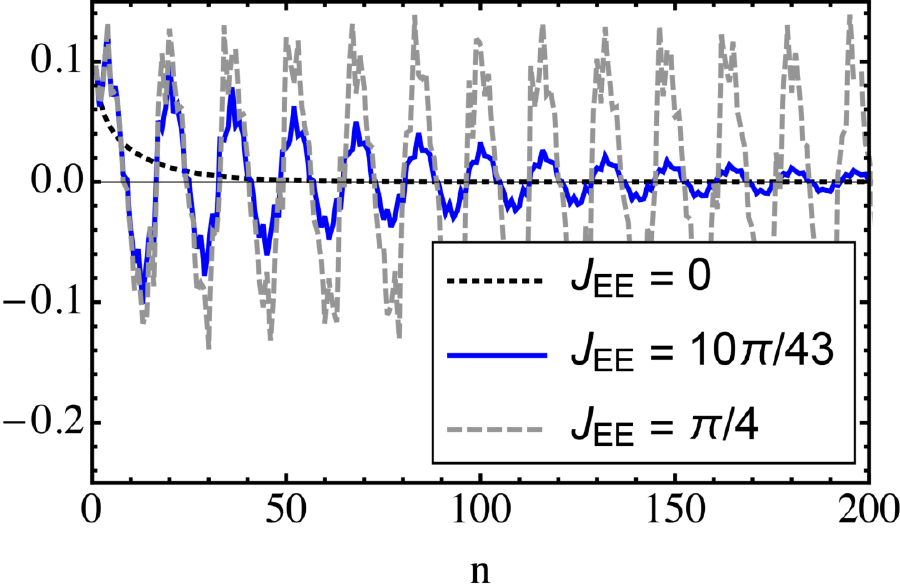}}
\quad
{\includegraphics[width=0.47\textwidth]{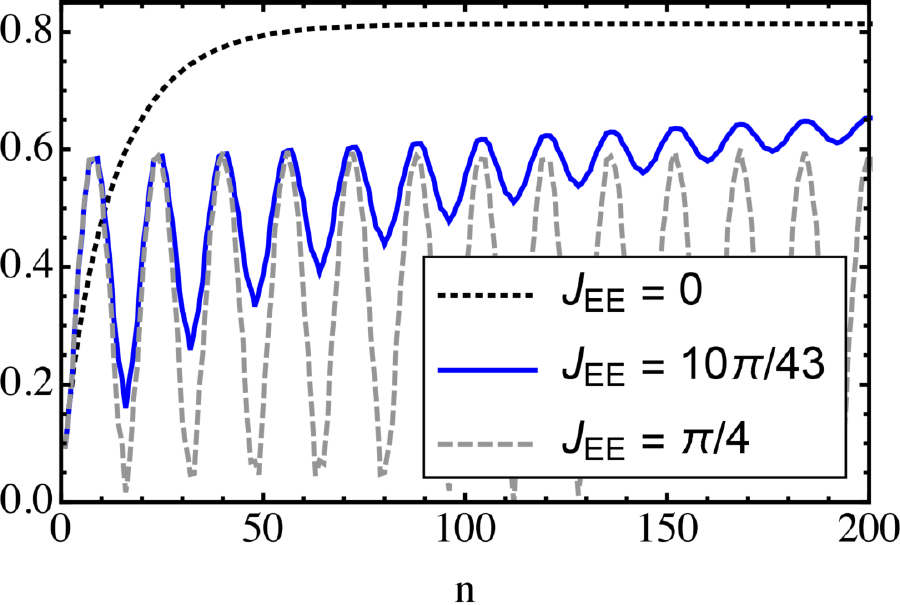}}
\caption{\textbf{Landauer bound} checked by plotting the difference $\beta \Delta Q - \Delta S$ [panel {\bf (a)}] between the instantaneous heat and entropy exchanges, and the difference $\beta Q - S$ [panel {\bf (b)}] between the respective cumulative quantities \ref{eq:CumulativeQS}, thus checking the bound from the beginning of the process up to step $n$. All quantities plotted against the number of collisions $n$. Whenever $\beta \Delta Q - \Delta S$ or $\beta Q - S$ become negative, the bound is violated. We chose the inter-environment coupling constant as $J_{EE}=0$ [black, dotted], $J_{EE}=10\pi/43$ [blue, continuous], and $J_{EE}=\pi/4$ [grey, dashed], for a Markovian, intermediate and strongly non-Markovian dynamics respectively. The other parameters used in these simulations are the same as in figure~\ref{fig:trace-dist}. In the Markovian case, both the instantaneous and the cumulative bound are always satisfied. Non-Markovianity in the dynamics however can cause the instantaneous bound to be violated, while the cumulative one is still always satisfied.}
\label{fig:landauer}
\end{figure}

Figure~\ref{fig:trace-dist} shows the trace distance between two different states evolving in time. In our case, the initial states that maximize the degree of non-Markovianity $\cn$, given by~(\ref{eq:N}), are $\ket{\pm}=(\ket{0} \pm \ket{1})/\sqrt{2}$, up to a global phase factor. Figure~\ref{fig:landauer} presents the behaviour of the Landauer bound formulated both in terms of the (discrete) flux $\beta \Delta Q$ and the change in entropy $\Delta S$, and of the cumulative quantities 
\beq
\label{eq:CumulativeQS}
Q_n = \sum_{l=1}^n \Delta Q_l, \qquad S_n = \sum_{l=1}^n \Delta S_l \equiv S(\rho^{\cs}_0) - S(\rho^{\cs}_n).
\eeq
The negativity of $\beta \Delta Q_n - \Delta S_n$ implies the violation of the principle, which occurs repeatedly in the non-Markovian case. Such violation is closely connected to the presence of correlations between the system and $E_n$ \emph{before} their direct interaction. The \emph{mutual information} 
\[
I(\rho_{\cs},\rho_{E} )=S(\rho_{\cs})+S(\rho_{E})-S(\rho_{\cs E})
\]
gives a measure of the correlations, whose presence is in explicit contradiction with one of the hypotheses behind the validity of the Landauer principle~\cite{ReebWolf14}. Nonetheless, the cumulative quantity $\beta Q_n - S_n$ remains positive at all times. Our analysis shows that the changes in entropy $\Delta S_n$ oscillate in time much more than $\beta \Delta Q_n$: such oscillations are responsible for the point-like violation of Landauer bound. 

\begin{figure}[p]
\center{{\bf (a)}\hskip7.5cm{\bf (b)}}\\
{\includegraphics[width=0.47\textwidth]{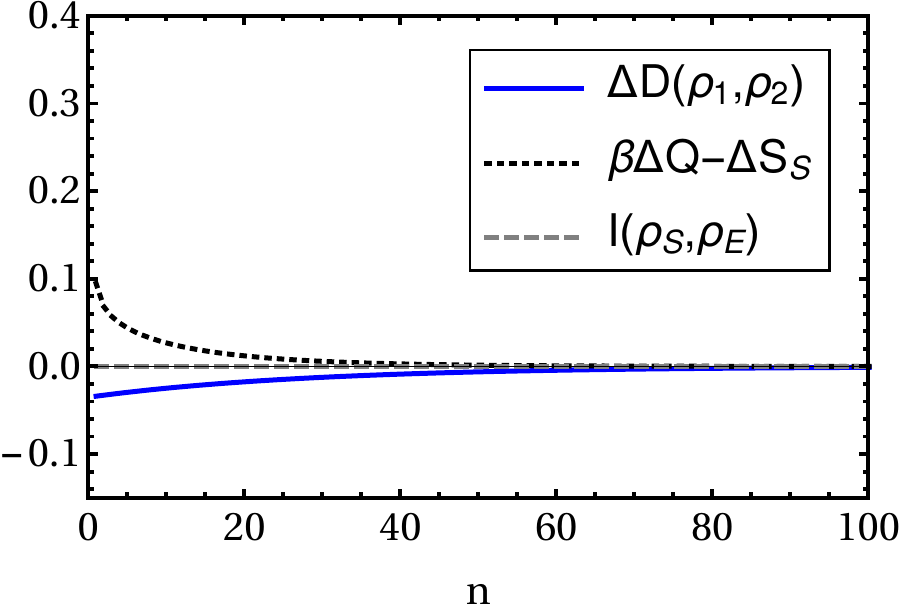}}
 \quad
{\includegraphics[width=0.48\textwidth]{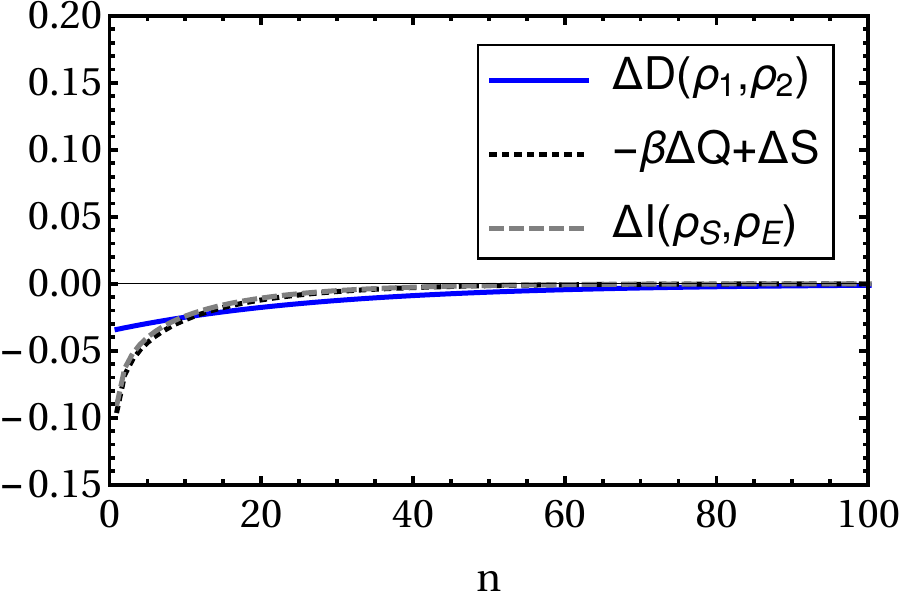}}
\center{{\bf (c)}\hskip7.5cm{\bf (d)}}\\
{\includegraphics[width=0.47\textwidth]{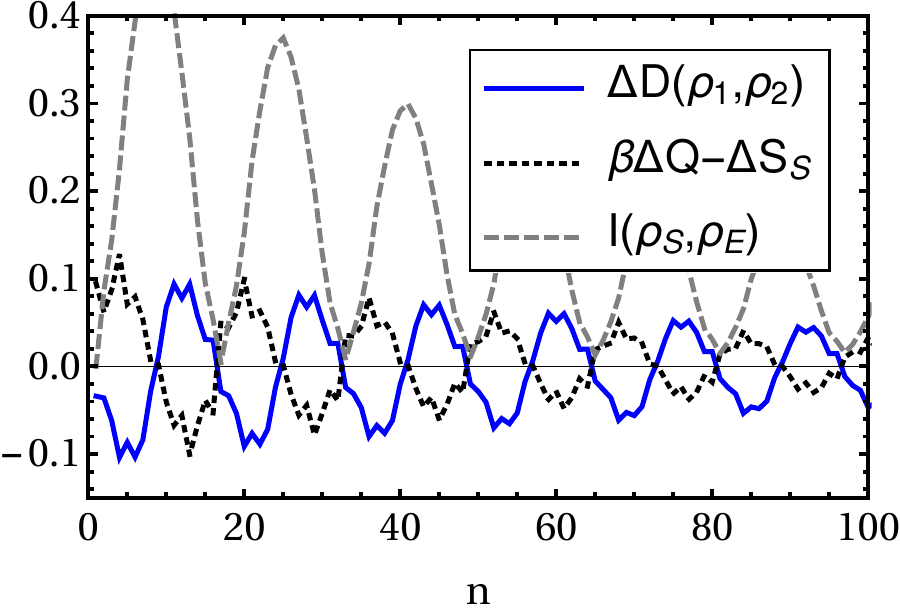}}
\quad
{\includegraphics[width=0.48\textwidth]{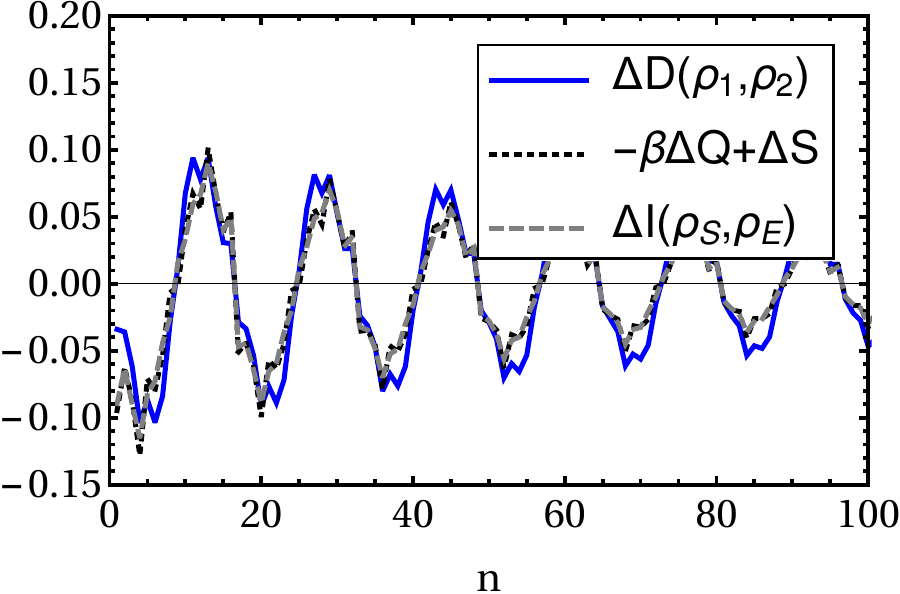}}
\center{{\bf (e)}\hskip7.5cm{\bf (f)}}\\
{\includegraphics[width=0.47\textwidth]{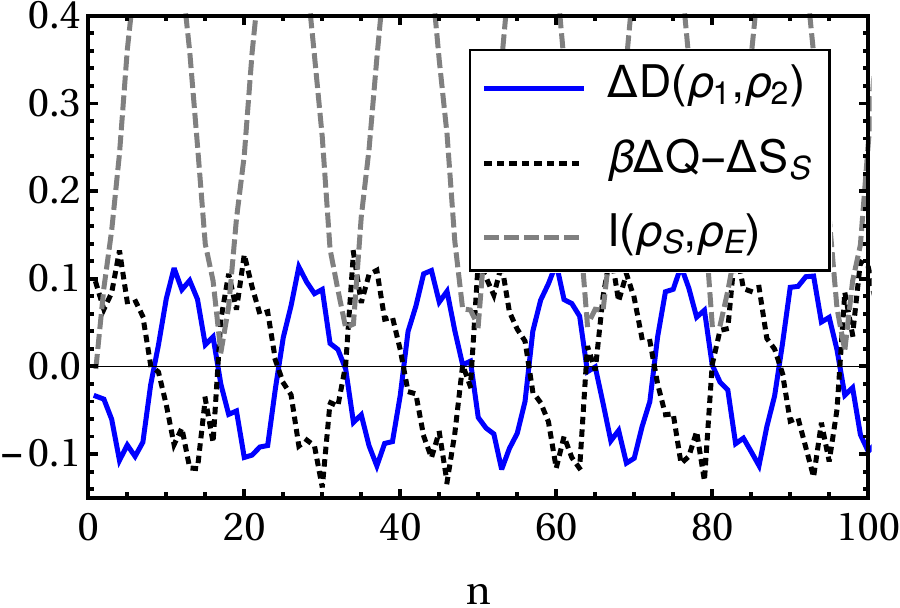}}
\quad
{\includegraphics[width=0.48\textwidth]{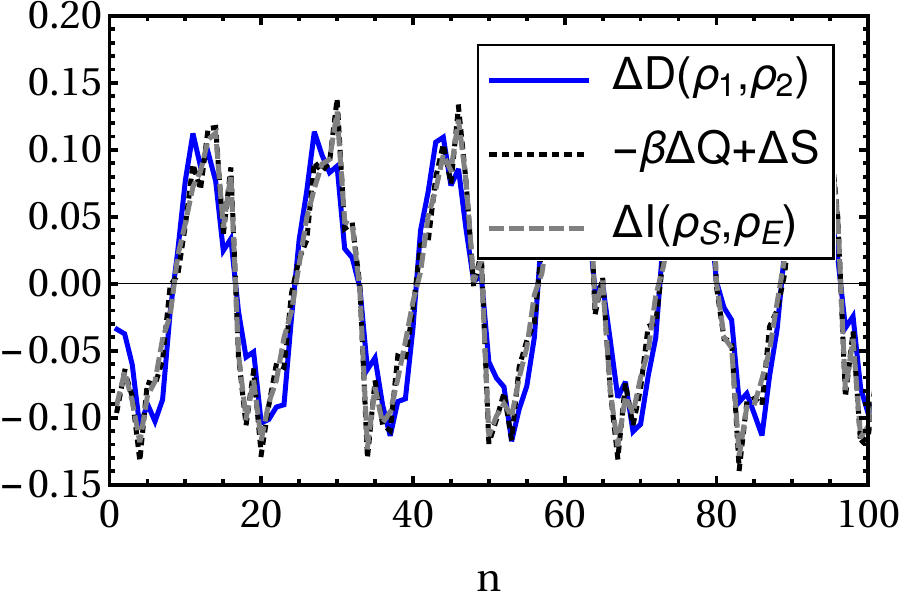}}
\caption{\textbf{Non-Markovianity, correlations and Landauer bound. Left panels: } in these plots non-Markovianity and correlations are represented respectively by the  derivative of $D(\rho_1,\rho_2)$ and by the mutual information $I(\rho_{\cs},\rho_E)$. The synchronous behaviour of the three plotted quantities is evident, hinting at their interconnectedness and common cause. {\bf Right panels:} in these plots the check on the Landauer bound is shown with reversed sign, $-\beta \Delta Q + \Delta S$, and we take the \emph{derivative} $\Delta I(\rho_{\cs},\rho_E)$ of the mutual information, to emphasize the connection between the three quantities examined. We have taken the inter-environment coupling constant $J_{EE}=0$ for {\bf (a)} and {\bf (b)}, $J_{EE}=10\pi/43$ for {\bf (c)} and {\bf (d)} and $J_{EE}=\pi/4$ for {\bf (e)} and {\bf (f)}, resulting in an increasingly non-Markovian dynamics. The parameters used in this simulations are as in figure~\ref{fig:trace-dist}.}
\label{fig:alltogether}
\end{figure}

Figure~\ref{fig:alltogether} summarizes the main results of this work. At the beginning the dynamics is Markovian, then the interactions build up correlations which grow strong enough to cause a shift to the non-Markovian regime. As the dynamics gets more and more non-Markovian and the trace distance increases, correlations diminish until they become negligible, at which point the dynamics returns within the Markovian regime, and the pattern repeats itself. The trace distance (discrete) derivative, the instantaneous Landauer bound and the (discrete) time derivative of mutual information proceed in a striking synchronous behaviour. The connection between $\beta \Delta Q - \Delta S$ and $\Delta I(\rho_{\cs},\rho_E)$ can be understood intuitively as both are well approximated by the derivative of the von Neumann entropy $S$, the leading contribution to both in this parameter regime. However, a deeper and more complete theoretical explanation of our findings is still missing.

\section{Concluding summary and remarks}
\label{conc}

We have studied the open-system dynamics undergone by a spin-1/2 particle through a sequence of discrete-time collisions with the elements of a spin environment. The asymptotic behaviour of the dynamics shows homogenization when the environmental particles are all in the same thermal state. This behaviour is maintained when the state of the environmental particles fluctuates weakly across their ensemble. For more significant fluctuations, however, homogenization is broken. 

By allowing for inter-environment interactions, we have introduced memory effects in the dynamics of the system, which shows features of non-Markovianity. We have investigated the connection between the emergence of such behaviour, the creation of system-environment correlations, and the observed instantaneous violations of the Landauer bound for the system particle. The feedback of excitations from the environment to the system enabled by the inter-environment interactions results in the building up of system-environment correlations and invalidates one of the assumptions of the quantum formulation of Landauer principle, thus causing its break down. We observed a striking synchronous behaviour between the instantaneous emergence of non-Markovianity in the dynamics, the establishment of system-environment correlations and the mentioned violations of the Landauer bound. The three interconnected behaviours are clearly originated by a common origin, which will be the subject of our forthcoming investigations.

\section*{Acknowledgments}

M Pezzutto thanks the Centre for Theoretical Atomic, Molecular, and Optical Physics, School of Mathematics and Physics, Queen's University Belfast for hospitality during the development and completion of this work. M Pezzutto and Y Omar thank the support from Funda\c{c}\~{a}o para a Ci\^{e}ncia e a Tecnologia (Portugal), namely through programmes PTDC/POPH/POCH and projects UID/EEA/50008/2013, IT/QuSim, IT/QuNet, ProQuNet, partially funded by EU FEDER, and from the EU FP7 project PAPETS (GA 323901). M Pezzutto acknowledges the support from the DP-PMI and FCT (Portugal) through scholarship SFRH/BD/52240/2013. M Paternostro acknowledges financial support from John Templeton Foundation (grant ID 43467), the EU Collaborative Project TherMiQ (Grant Agreement 618074), and the Julian Schwinger Foundation (grant number JSF-14-7-0000). All authors gratefully acknowledge support from the COST Action MP1209 "Thermodynamics in the quantum regime".

\end{document}